\documentclass[smallextended]{svjour3}       
\smartqed  
\usepackage{graphicx}
\usepackage{amsmath}
\usepackage{bm}
\usepackage{authblk}
\begin{document}

\title{
Compressibility, laws of nature, initial conditions and complexity}

\author{Sergio Chibbaro
\and Angelo Vulpiani}

\institute{S. Chibbaro \at
              Sorbonne Universit\'e, UPMC Univ Paris 06, Institut Jean Le Rond d'Alembert, F-75005, Paris, France and CNRS, UMR 7190,  F-75005, Paris, France     
           \and
           A. Vulpiani \at
              Dipartimento di Fisica, Universit\`a "La Sapienza", Piazzale Aldo Moro 2, I-00185 Roma, Italy
and  Centro Interdisciplinare B. Segre, Accademia dei Lincei, Roma, Italy
}

\date{Received: date / Accepted: date}

 \maketitle
 
\begin{abstract}
We critically analyse the point of view for which laws  of nature 
are just a mean to compress data. 
Discussing some basic notions of dynamical systems and information theory, we show that 
the idea that  the analysis of large amount of data by means of an algorithm of compression is
equivalent to the knowledge one can have from scientific laws, is rather naive.
In particular  we discuss the subtle conceptual topic of   
the initial conditions of phenomena which are generally incompressible.
 Starting from this point, we argue that laws of nature represent more than a pure compression of data, and that the availability of large amount of data, in general,  is not particularly useful  to understand the behaviour of complex phenomena.
 \end{abstract}
 
\section{Introduction}
It is  not necessary to stress too much the   fact
that the external world is not just a jungle of irregular events.
There is a quite clear  evidence of our ability  to understand (at least partially)  the 
many regularities of our physical world.
Then, it is  quite natural to ponder about the origin of such a success.
A very general question is: why is the physical world comprehensible? 
In particular one can wonder about the existence and the status of 
mathematical laws which allow us \emph{quantitative or qualitative} predictions in agreement with experiments\cite{seve3,bailly2006mathematiques}. 

In the past some  scientists (and philosophers)  stated that the aim of science is to organise in
the most economical fashion the data collected from experiments.
{In this view, laws  are just a very effective way to compress disparate data.
}Likely the most important champion of such a view of the science  has been  E.Mach\cite{Mach1,Mach2}:

 \noindent
 {\it  The so-called descriptive sciences must chiefly remain content with reconstructing
individual facts . . . But in sciences that are more highly developed, rules for the
reconstruction of great numbers of facts may be embodied in a single expression.

Thus, instead of noting individual cases of light-refraction, we can mentally reconstruct
all present and future cases, if we know that the incident ray, the refracted
ray, and the perpendicular lie in the same plane and that $\sin \alpha/ \sin \beta = n$. Here,
instead of the numberless cases of refraction in different combinations of matter
and under all different angles of incidence, we have simply to note the rule above
stated and the values of n, which is much easier. The economical purpose is here
unmistakable.}

This  point of view has been  shared by many scientists of the positivism or neopositivism currents.
Interestingly,  such an approach
has been recently reconsidered in the framework of algorithmic
complexity \cite{li2009introduction} by researchers without specific philosophical interests.
 For instance, Solomonoff, one of the
fathers of the theory, considers (without any reference to Mach) a scientific law, and
more generally a theory, as an algorithm for compressing the results of experiments,
providing a mathematical formalisation of the idea of science as an economy of
thought \cite{solomonoff1964formal}:

\noindent
{\it The laws of science that have been discovered can be viewed as summaries of large amounts
of empirical data about the universe. In the present context, each such law can be transformed
into a method of compactly coding the empirical data that gave rise to the law.}

\noindent We can cite other similar opinions, e.g.
 \\
{\it The existence of regularities may be expressed by saying that the world is algorithmically
compressible. Given some data set, the job of the scientist is to find a
suitable compression, which expresses the causal linkages involved. For example,
the positions of the planets in the solar system over some interval constitute a
compressible data set, because Newton's laws may be used to link these positions
at all times to the positions (and velocities) at some initial time. In this case
NewtonÕs laws supply the necessary algorithm to achieve the compression.}
\cite{davies1990physical}.
\\
\\
{\it The intelligibility of the world amounts to the fact that we find it to be algorithmically
compressible. We can replace sequences of facts and observational data
by abbreviated statements which contain the same information content. These
abbreviations we often call laws of Nature. If the world were not algorithmically
compressible, then there would exist no simple laws of Nature.}\cite{barrow2007new}.

As an interesting exception to the idea of science as  economy of thought, we may recall Born~\cite{born1949natural} who ironically noted: 
 
 {\it if we want to
economise thinking, the best way would be to stop thinking at all, and then the expression {\it economy
of thinking} may have an appeal to engineers or others interested in practical applications, but
hardly to those who enjoy thinking for no other purpose than clarify a problem}.

In our opinion the idea of economy of thought  removes all objectivity to the scientific laws and mathematical constructions.
For reasons which we do not discuss here, this approach has gained 
much interest and broad success in the last decades under the vaguely-defined concept of complexity. 
The word ``complexity" has become rather a ``logo" for mainstream analysis;
like in the 30's to be ``modern" was mandatory, to be ``complex" seems required to be fashionable today
 ~\cite{Sok_97}.
While we can agree that the arising of nonlinear physics and mathematics since the pioneering works of Poincar\'e has represented a major change in science, perhaps even a change of paradigm, the recent insistence on the ``complex" often appears
  preposterous~\cite{ccv}.
Loosely speaking, complexity studies share the idea to apply the same tools, 
mainly  from dynamical systems and statistical mechanics,  
to a very large spectrum of phenomena, from social and human sciences to astrophysics, regardless of the specific content of each problem. Therefore it is implicitly assumed that the laws underlying these phenomena are not important.
More recently, this point of view has become extremely radical with the "big data" philosophy, which presents many
conceptual and technical problems~\cite{ccv,coveney2016big,crutchfield2014dreams}.
In that framework, laws should be supplanted by the statistical analysis of a large amount of data, again carelessly of any specificity. In this sense, science becomes
a technical compression of data. 

The  point of view of science as economy of thought  seems to be in agreement with the idea that
   the central goal of science has been thought to be "prediction and control".
 As a relevant example of this opinion  
 we  can mention  the von Neumann's
belief that powerful computers and a clever use of numerical analysis would eventually
lead to accurate forecasts, and even to the control of weather and climate:

{\it The computer will enable us to divide the atmosphere at any moment into stable regions and
unstable regions. Stable regions we can predict. Unstable regions we can control.}\footnote{Cited
by Dyson\cite{dyson2009birds}.}

We know now that the great  von Neumann was wrong:   he did not take into account   the role
of  chaos.
About half a century ago, thanks to the contribution of M.H\'enon, E.Lorenz and B.V.Chirikov (to cite just some of the most eminent scientists in this field),
we had the (re)discovery of deterministic chaos.
This event was scientifically important, e.g. to clarify topics as the different possible origin of the statistical laws and the intrinsic practical
limits of  predictions.
Yet one has to admit that the topic of  "chaos" 
induced  a certain confusion
about concepts as determinism, predictability, stochastic laws and the understanding of a phenomenon in terms of compression;
for instance, Davies \cite{dyson2009birds} writes
 
{\it 
there is a wide class of physical systems, the so-called chaotic ones, which are not algorithmically
compressible.}
\\
We will see  that
 what is not compressible is the time sequence generated by chaotic systems, and
this is due to the non-compressibility of a generic initial condition.
\\
The aim of the present paper is  to clarify the real relevance of the ideas as ''compression'' and algorithmic complexity
in science, in particular in physics.
For such an aim,  a detailed analysis of of laws, initial conditions and data is necessary.

Section 2 is devoted to a  general discussion on evolution laws, initial conditions and data.
In Section 3 we treat in details the role of the initial conditions and the compressibility.
The question of the relation between compression and laws of natures is discussed
in Section 4. 
In order to clarify the role of the algorithmic complexity in the research, 
two case studies are presented in Section 5.
Some final remarks and considerations in Section 6.

\section{About the laws and data}

The  aim of this  Section is to clarify the distinction between  
phenomena, evolution laws, initial conditions, and series of data;
the level of the discussion will not  be  technical, the precise notion of complexity
will be discussed in Section 3 and in the Appendix.

Natural phenomena, roughly speaking, can be divided in two large classes:
\begin{enumerate}
\item[($a_1$)] simple,  e.g. stable  and predictable, an example is the pendulum;
\item[($a_2$)]  complex, i.e. irregular and unpredictable, a paradigmatic example is turbulence in fluids.
\end{enumerate}
The evolution laws  can be:
\begin{enumerate}
\item[($b_1$)] known;

\item[($b_2$)]  unknown.
\end{enumerate}
The initial conditions can be:
\begin{enumerate}
\item[($c_1$)] simple;

\item[($c_2$)]  complex.
\end{enumerate}
Finally, the series of data generated by a certain phenomenon appear:
\begin{enumerate}
\item[($d_1$)] regular;

\item[($d_2$)]  irregular.
\end{enumerate}

 \subsection{The simplest case: the law is known}
 First, let us consider  the case $b_1$, which is, from a methodological point of view, the simplest one.
 The possibilities are the following:
$$
 a_1\, + \, c_1 
 \eqno(S.1)
 $$
 $$
 a_1\, + \, c_2 
 \eqno(S.2)
 $$
 $$
 a_2\, + \, c_1 
 \eqno(C.1)
 $$
 $$
 a_2\, + \, c_2 
 \eqno(C.2)
 $$
The cases S.1 and S.2 are quite clear:
independently from  initial conditions, the system will display a regular behaviour, thus 
$$a_1\, + \, c_1 \to \,  d_1, ~~{\text and}~~  a_1\, + \, c_2 \to \,  d_1~.$$

The case C.1 is not typical (i.e. rather rare);
 on the contrary the case C.2  must be considered \textit{generic},
i.e.  considering a large ensemble of natural phenomena, almost all of them will join this class. In turn, ``almost all'' indicates that the probability to find one behaving differently will be basically zero if we consider a large enough ensemble.
Indeed the situation C.1 can be observed only under particular circumstances, most of natural phenomena are part of the category C.2. 
Why initial conditions are almost always complex
will be explained  in Section 3.

Considering now the case C.2, since the  evolution law of the  system is  known, 
 the irregularity in the outcome of the dynamics has to be necessarily hidden in the initial conditions. 
To better clarify this point, one can consider that any initial condition  $x_0$ can be written, in a unique way,  in terms of a binary sequence
 $\{ i_1, i_2, ... \}$. 
 Such a sequence can be
 \begin{description}
\item[ $\star$ ] compressible (for instance periodic, or periodic after a certain initial part), corresponding to rational $x_0$;
in such a case the dynamics generates a regular sequence,  as in the C.1 case:
$$a_2\, + \, c_1 \to \,  d_1~.$$
\item[$\star$] incompressible (e.g. aperiodic),  
in such a case the dynamics generates an irregular sequence,  as in the C.2 case:
$$a_2\, + \, c_2 \to \,  d_2~.$$
\end{description}

\subsection{The evolution law is not known}
Of course  only in few lucky situations (mainly in physics)
we know the laws ruling a certain phenomena with a good precision.
In ecology, biology  and many other applied sciences, 
 it is not possible  to write down the equations describing  a certain phenomenon on the basis
of well established theoretical frameworks and it is unavoidable to use
a combination of intuition and experimental data~\cite{chibbaro2014reductionism,times}.

Often the importance of the  concept of {\it state 
of the system}, i.e. in mathematical terms, the variables which describe the phenomenon 
under investigation is not enough stressed.
The relevance of this aspect is usually underestimated;
only in few cases, e.g. in mechanical systems, it is easy to identify
the variables which describe a given phenomenon.
On the contrary, in a generic case,  there are serious difficulties;
we can say  that the main effort in building a theory on nontrivial phenomena  concerns 
the identification of the appropriate variables.
Such a difficulty is well known in physics, for instance 
 in the context of  statistical mechanics  Onsager and
Machlup, in their seminal work on fluctuations and irreversible processes~\cite{onsager1953fluctuations},
stressed the problem  with the caveat:
{\it how do you know you have taken enough variables, for it to be Markovian? }
\footnote{
As  paradigmatic example  let us consider the Langevin equation 
$$
{d^2 x \over dt^2}+ \gamma {dx \over dt}=-\omega^2 x +c \eta
$$
where $\eta$ is a white noise, i.e. a Gaussian stochastic process with $\langle \eta \rangle=0$ and
$\langle \eta(t) \eta(t') \rangle = \delta (t-t')$,  and $\gamma>0$.
It is worth emphasising that the vector ${\bf y}=(x, dx/dt)$ is a Markov process, i.e. its stochastic evolution
at $t>0$ is determined only by ${\bf y}(0)$, on the contrary the scalar variable $x$ is not a Markovian process, and thus its dynamics depends on its past history.
}

In a similar way, Ma notes that\cite{Ma}:
{\it  the hidden worry of thermodynamics is: we do not know how many coordinates
or forces are necessary to completely specify an equilibrium state.}

Usually we have no definitive method for selecting  the proper
variables and only a deep theoretical understanding can suggest the ``good ones''.

Takens~\cite{takens1981detecting} showed that from the study of a time series $\{ u_1, u_2, ..., u_M \}$, where $u$ is an observable,
it is possible (if we know that the system is deterministic and is described by a finite dimensional vector) to determine
a proper set of variables fully describing the system .
Unfortunately, in practice,  the method has rather severe limitations:
\begin{description}
\item[ (a)]  it works only if we know a priori that the system is deterministic;
\item[ (b)]  because of the finite size $M$ of the time series, in practice the protocol fails if the dimension of the attractor is large enough (say more than $5$ or $6$). 
\end{description}
Therefore this method cannot be used, apart from special cases (with a small dimension), to build  a model
for the evolution law from data\cite{eckmann1992fundamental}.

We will discuss in some details the difficulties of the method in Section 5.2.

\section{Evolution laws and initial conditions}
Let us now consider  a topic  which should be carefully
analyzed: the role of initial conditions, which are usually independent of the laws
of nature. Such an important point had been already realised by Newton\cite{newton1934sir}
who noted that all the planets move in the same direction on concentric orbits, while
the comets move in eccentric orbits, concluding that such a property of the solar
system cannot be a mere coincidence, but it is due to the initial condition.
Wigner  considers the understanding of the distinction between laws and initial conditions
as the most important contribution that Newton made to science, even more
important than the laws of gravitation and dynamics\cite{wigner1963problem}.

\subsection{Two examples of deterministic systems}

Let us analyse the very different behaviour of two deterministic systems;
we will see how the initial conditions can play a basic role.
\\
{\bf Example A} The pendulum of length $L$:
\begin{equation}
{ d^2 \theta \over dt^2}= -{g \over L}\sin \theta \,\,.
\end{equation}
From  well known mathematical theorems  on differential equations 
we know that:
\\
a)  the initial condition $(\theta(0), d\theta(0)/dt)$ determines in a unique way the state of the system
$(\theta(t), d\theta(t)/dt)$ at any time $t$;
\\
b) the motion is periodic, i.e. there exists a time $T$ (depending on the physical parameters)
 such that  
$$
\Big( \theta(t+T), {d\theta(t+T) \over dt} \Big)=\Big(\theta(t), {d\theta(t) \over dt} \Big) \,\, ;
$$
\\
c) the  time evolution  can be expressed in terms of a 
function of $t$ and the initial conditions:
$$
\theta(t)= F\Big(t, \theta(0), {d\theta(0) \over dt} \Big)\,\, .
$$
The function $F$ can be written in an explicitly way  if 
 $\theta(0)$ and  $d\theta(0)/dt$ are small, in such  case $T=  2 \pi \sqrt{L /g}$:
 $$
 \theta(t)= \theta(0) \cos (\omega t) + {1 \over \omega}{d \theta(0) \over dt} \sin(\omega t)
 $$$$
 {d \theta(t) \over dt}= -\theta(0) \omega  \sin (\omega t) + {d \theta(0) \over dt} \cos(\omega t)
 $$
 where $\omega= 2 \pi/T$.
 In the generic case, $F$ can be easily determined with the wished precision by numerical resolution.
 \\
 \\
{\bf Example B} The Bernoulli's shift:
\begin{equation}
x_{t+1}= 2 x_t \,\,\, mod \,\, 1 \,.
\label{eq:2}
\end{equation}
Where the operation $mod\,\, 1$ corresponds to  take  the fractional part of a number, e.g. $1.473 \,\, mod\,\, 1=0.473$.

It is possible to show that the  Bernoulli's shift  is chaotic: 
a small error in the initial conditions doubles at every step. 
Consider an initial condition  $x_0$  in the interval $[0,1]$, it can be expressed by an infinite sequence
of $0$ and $1$:
\begin{equation}
x_0={a_1\over 2}+{a_2\over 4}+ ... +{a_n\over 2^n}+ ...
\end{equation}
where every $a_n$ takes either the value $0$ or the value $1$. 
The above binary notation allows us to determine the time evolution by means of
a very simple rule: at every step, one has just  to move the "binary point" of the binary expansion of
$x_0$  by one position to the right and eliminate the integer part.
 For example, from
 $$
 x_0=0.11010000101110101010101100.....
 $$
 one has
$$
 x_1=0.1010000101110101010101100.......
 $$
$$
 x_2=0.010000101110101010101100.........
 $$
$$
 x_3=0.10000101110101010101100...........
 $$
 and so on. In terms of the sequence$\{ a_1, a_2, ....\}$ it becomes quite clear how
crucially the temporal evolution depends on the initial condition.
 If the binary sequence associated to $x_0$ is not complex, e.g. $x_0$
 is a rational number, then the sequence $\{ x_0, x_1, ...\}$ will be regular;
 on the contrary if the binary sequence associated to $x_0$ is  complex,
  the  sequence $\{ x_0, x_1, ...\}$ will remain complex.
   
Let us consider two initial conditions $x^{(1)}_0$ and $x^{(2)}_0$ such that
$|x^{(1)}_0-x^{(2)}_0| < 2^{-M}$ for some arbitrary (large) integer number $M$,
 this means that  $x^{(1)}_0$ and $x^{(2)}_0$   have
the first $M$  binary digits identical, and they may differ only afterwards. 
The above discussion shows that the distance between the points increases rapidly:
for $t <M$ one has an exponential growth of the distance between the two trajectories
 \begin{equation}
|x^{(1)}_t- x^{(2)}_t |\sim |x^{(1)}_0- x^{(2)}_0|\, 2^t \,\, ,
\end{equation}
As soon as $t>M$ one can only conclude that
 $|x^{(1)}_t- x^{(2)}_t|< 1$.
 We can say that 
 our system is chaotic: even an arbitrarily small error in the initial conditions eventually
dominates the solution of the system, making long-term prediction impossible.

From the above brief discussion, we see how  in deterministic systems one can have 
the following possible cases (in decreasing order of predictability):
\begin{enumerate}
\item Explicit possibility to determine the future (pendulum in the limit of small oscillations);
\item Good control of the prediction, without an explicit solution (the pendulum with large oscillations);
\item  Chaos and practical impossibility of predictability  (Bernoulli's shift).

\end{enumerate}

\subsection{About initial conditions and compression}

Let us consider again the   dynamical system (\ref{eq:2}), which
is chaotic, i.e. the distance between two trajectories initially very close, increases exponentially in time:
\begin{equation}
\delta_t= |x^{(1)}_t- x^{(2)}_t |\sim \delta_0 \, e^{\lambda t} \,\, ,
\end{equation}
where $\lambda$, called Lyapunov exponent, is positive.
In the  system (\ref{eq:2}),  $\lambda= \ln 2$,
this means that a small error in the initial conditions doubles at every step. 
Suppose that $x_0$ is a real number in the interval $[0,1]$, it can be expressed by an infinite sequence
of $0$ and $1$ as in (3).
We already saw that looking at  the sequence
$\{ a_1, a_2, .... , a_n, ... \}$,
 it becomes quite clear how
crucially the temporal evolution depends on the initial condition.

Let us now make a brief digression on the notion of ''complexity'' of a binary
sequence. Generally speaking, different types of sequences are possible, for example
consider the following ones:
\begin{equation}
 11111111111111...
\end{equation}
\begin{equation}
 10101010101010...
\end{equation}
\begin{equation}
 00101000110100...
\end{equation}
 It is quite natural to say    that sequences (6) and (7) appear to be ''ordered'',
whereas sequence (8) seems ''complex''. Why should one classify the sequences
in this way?
In the case of (6) and (7) the knowledge of the first $n$  values
 $\{ a_1, ..., a_n\}$
appears to be sufficient to predict the following values 
$\{ a_{n+1}, a_{n+2}, ...\}$.
This is not
true for sequence (8), which seems to be generated by a stochastic, rather than
 deterministic rule. In this case, one could think that the sequence of $0$ and $1$
is generated tossing a coin, and writing $1$ for heads and $0$ for tails. One way to
formalise this intuitive concept of complex behaviour is to associate it with the
lack of a constructive rule; then the cases  (6) and (7) are not complex because
they can be generated by means of very simple rules. On a computer, for instance,
(6) can be generated through a single statement:
$$
{\tt  WRITE \,\,\, 1  \,\,\, N \,\,\, TIMES}
$$
and similarly for (7):
$$
{\tt WRITE \,\,\, 1 0 \,\,\, N/2 \,\,\, TIMES}
$$
By contrast, (8) seems to require a program of the kind:
$$
{\tt WRITE \,\,\, 0 \,\,\, 
WRITE \,\,\, 0 \,\,\,
WRITE \,\,\, 1 \,\,\,
WRITE \,\,\, 0 \,\,\, ......}
$$
We can conclude that the sequences (6) and (7) can be considered ``simple''
because they can be obtained with a short computer code; on the contrary the length of 
computer code which generates (8) is proportional to the size of the sequence.

The rationalization  of the above remarks needs the introduction of 
a precise mathematical formalisation of the algorithmic complexity of a sequence\cite{li1992inductive,li1994kolmogorov,ming2014kolmogorov}, a brief introduction is given in the Appendix.

\section{Do scientific laws compress empirical data?}

Surely  it is fair to say that once a scientific law has been established
one has a sort of compression, however such a conclusion deserves a careful analysis.
In order to clarify such a topic we briefly discuss two  examples of
scientific laws, namely the Newton's equations for the classical mechanics and the
Schr\"odinger equation.

\subsection{Classical mechanics and astronomy}
It is well know that from Newton's equations
and the gravitation law, one can derive many important astronomical facts,
for instance  Kepler's laws.

On the other hand it is not completely  correct to conclude that   Newton's equations
and the gravitation law are able to compress all the astronomical behaviors.
After the seminal contribution of Poincar\'e, we know that a system of three bodies 
interacting with the gravitational force   is usually chaotic~\cite{ccv}.
Such a celebrated system  is  an example of the case (C.2).

In the following we will show how the presence of chaos implies the failure of the possibility
to compress astronomical evolution.
For sake of simplicity, instead of discussing such a difficult problem, we  reconsider  the system (2)
which shares many features with the three body problem.

Let us analyse the problem of  transmission to a friend, with accuracy $\Delta$, of a sequence
$x_t \, 0<t<T$, generated by the rule (2).
 At first glance, the problem seems quite simple:
we could opt for transmitting $x_0$
and the rule (2), which costs a number of bits independent of $T$.
 Our friend  would then be left with the task of generating the sequence 
$x_1, x_2, ... , x_T$.
However, we must also choose the number of bits to which $x_0$ should be specified.
From (3), the accuracy $\Delta$ at time $T$ requires accuracy 
$\delta_0 \sim 2^{-T} \Delta$  for $x_0$, hence
 the number of bits specifying $x_0$ grows with $T$. Again, we have to tackle the
problem of the complexity of a sequence of symbols, 
$\{ a_0, a_1, ... \}$.
 The fact is that there are ''simple'' initial conditions, of the type (6) or (7), which can be specified
by a number of instructions independent of the length of the sequence, but there are
complex sequences as well.

We saw that the evolution law of (2) is nothing but a shift of the binary point of
the sequence $\{ a_1, a _2, ..., \}$.
Therefore we have that the evolution of $x_0$ is regular (e.g. periodic) if its sequence
 $\{ a_1, a _2, ..., \}$ is not complex while it is irregular if
 $\{ a_1, a_2, ..., \}$ cannot be compressed.
 
 So we have that both in systems with regular behavior (the pendulum) and
 chaos (the Bernoulli's shift), it is straightforward to compress the evolution law.
 The difference between the  two systems is in the output which is
 always regular in the pendulum, whereas in the Bernoulli's shift it can be regular
 or irregular depending on the initial condition.
 
The  conclusion, somehow rather intuitive, is that 
in deterministic systems the
details of the time evolution are well hidden in the initial condition which turns out to be typically complex.
The complexity of initial conditions follows from an important mathematical  result of Martin-L\"of \cite{martin1966definition} who showed  that almost all infinite binary sequences, which express
the real numbers in $[0,1]$, are complex.
We do not enter into details of such a topic which involves rather subtle points related to the infinity
and the G\"odel theorem~\cite{caludelongo}.

Coming back to astronomy, from the previous result we can conclude that, in  presence of chaos,
the knowledge of the basic laws ruling the time evolution of the astronomical bodies
(i.e.  Newton's laws and the gravitational force) is not enough to compress the complex time behaviour
which is hidden in the (almost surely) complex initial condition.

\subsection{Quantum mechanics and chemistry}
Consider now the Schr\"odinger equation and its relation with chemistry;
P.A.M. Dirac wrote the following celebrated sentence~\cite{dirac1929quantum}:

{\it The fundamental laws necessary for the mathematical treatment of a large part of physics and the whole of chemistry are thus completely known, and the difficulty lies only in the fact that application of these laws leads to equations that are too complex to be solved.}

We are able  to find the explicit solution of the Schr\"odinger equation for the hydrogen atom,
and such a result has been the starting point to explain with high accuracy the phenomena observed in experimental spectroscopy. 
 So at first glance it seems fair to say that the Schr\"odinger equation is able to compress the spectroscopic data.
 
 On the other hand the relation between quantum mechanics and chemistry is
 rather controversial, and surely
  much weaker than the link between Newtonian mechanics and astronomy. 
 For instance, in the latter case the theory was able to predict the existence of a previously unknown planet 
 (Neptune). On the contrary, as far as we know, there is nothing similar in chemistry regarding the prediction of a new element solely on the basis of quantum mechanics.
 
 Let us briefly discuss an issue  which allows us to understand 
the severe  limitation of the predictive power of quantum mechanics~\cite{chibbaro2014reductionism,Pri_81,Scerri2008}.
Consider the  pyramidal molecules, e.g. ammonia ($N H_3$), phosphine ($P N_3$) 
 or arsine ($As H_3$).
 The three isolated molecules are described by the same Hamiltonian
 with the unique numerical  difference of a parameter (namely the masses of the different chemical species $N$,
 $P$ and $As$).
 From an analysis  of  the quantum problem of the isolated molecule 
 one obtains that the pyramidal molecules are delocalised, in
 clear disagreement with  experiments which show that
 arsine is localised~\cite{Pri_81,Jon_10}.
 The localization does not follow in a straightforward way from quantum mechanics but is a consequence of the interaction of single molecules with an external environment consisting of a very large number of components.
 The emergence of molecular structures  can be understood only considering
 the   interaction with an environment containing a large number of microscopic constituents
  ~\cite{Pri_81,Jon_10}. 
{ In this case, we could say that it is not fair to speak of compression because of the (very) complex boundary conditions to be supplied to Schr\"odinger equations.
}
\section{About the role of algorithmic complexity: two case study}

In order to clarify  the role of chaos, initial conditions and 
algorithmic complexity in real scientific activity we discuss in some details two important
topics.
 Namely we consider the  features of fully developed turbulence (FDT)\cite{frisch1995turbulence},
and   how   to reconstruct the evolution law from
 time series in the cases it is not possible to use 
 some well established theory.

\subsection{Turbulence}
Turbulent flows, a paradigmatic case of complex system, are governed 
by the Navier-Stokes equations (NSE) which can be written in one line.
In the incompressible case one has:
$$
 \partial_t {\bf u}+({\bf u} \cdot \nabla) {\bf u}= -\frac{1}{\rho} \nabla p + 
 \nu \Delta {\bf u} + {\bf f} \,\, , \,\, \nabla \cdot {\bf u}=0 \,\, ,
$$
  where ${\bf u}$ is  the velocity field,  $\rho$  the (constant) density, $p$ the pressure,
  $\nu$ the kinematic viscosity and ${\bf f}$ an external force~\cite{frisch1995turbulence,bohr2005dynamical}.
 
So, naively, one could conclude  that, since we know the equation for the time evolution
of the velocity field,  somehow, the phenomenon of turbulence has been compressed, {as well as most of fluid mechanics}.
The study of   some specific  aspects  allows for  the understanding  
of the precise meaning and limitation of such a conclusion.
First, let us consider the   problem of the initial conditions:
of course in any experiment they  are necessarily known
with a limited precision.
A rather severe limitation is due to the fact that
in the limit of very large Reynolds numbers $R_e$
\footnote{
The Reynolds number
$$
R_e={U L \over \nu}~,
$$
\\
being $U$ and $L$ the typical velocity and length of the flow respectively, 
 indicates the relevance of the non linear terms.
At small $R_e$ we have a laminar flow, while the regime $R_e \gg 1$ is called fully
developed turbulence.}, 
for a proper description of the turbulent velocity field  
it is necessary to consider a huge  number of degrees of freedom: a rough estimate is ${\cal N} \sim R_e^{9/4}$~\cite{frisch1995turbulence,bohr2005dynamical}.
Therefore for the typical values of $R_e$ in FDT ($ \sim 10^6 - 10^9$),
because of  the gigantic amount  of data necessary to describe  the involved degrees of freedom,
we have an obvious impossibility to access to the initial conditions with the  proper  accuracy.

In addition at large $R_e$ the NSE are chaotic: the distance between two initially close initial conditions 
increases very fast.
Therefore, as a consequence of the practical impossibility to access  the initial conditions 
with high accuracy, and the presence of deterministic chaos,  even with a very powerful computer and
 accurate numerical algorithms,  it is not possible  to perform a simulation of 
 the NSE  for a long term and compare the single-trajectory prediction with experimental results.

Because of  the practical impossibility to compare the experimental results
with the numerical computation of the field ${\bf u}({\bf x}, t)$,
 we cannot say that the NSE are able to compress the turbulent behaviors.
Nevertheless there is  a general consensus on the validity of the NSE  for the  FDT.

We can mention at least four items supporting the opinion that NSE are able to describe FDT,
the agreement of the results observed in FDT and those obtained by the  NSE  for:
\begin{description}
\item[ (a)] short time prediction of the velocity field;
\item[ (b)] long time prediction of averaged (e.g. spatially caorse grained) quantities;
\item[ (c)] the scaling laws, and more generally, the statistical features;
\item[ (d)] the qualitative and quantitative spatio-temporal features (e.g. large scale coherent structures).
\end{description}
A general discussion can be found in the literature~\cite{frisch1995turbulence,bohr2005dynamical}.
 
\subsection{When the evolution law is not known}

Let us  note that the NSE have been derived
on a theoretical basis  using the Newton equations,  assuming the
hypothesis of the continuity of matter, and  some thermodynamic considerations.
One can wonder about the possibility to obtain
the  NSE just looking directly at experimental data.

Since in the NSE one deals with fields (i.e. infinite dimensional quantities),
it is natural to expect formidable difficulties.
A less ambitious (but conceptual similar) task
is   to  build models in finite dimension on the basis of experimental data~\cite{Kan_97}.
 Only for the sake of simplicity we assume the  most favorable case, i.e.
 the time is discrete,  the system is deterministic and  we know that  the state of the system  at time $k$ 
 is a finite dimensional vector ${\bf x}_k$.
 
Consider the problem of the prediction   from the available data, i.e. a long time sequence.
A quite natural approach is
  to search for a past state similar to the present state of a given phenomenon of interest,
 then, looking at the sequence of events that followed the past state, one may infer
by analogy the evolution that will follow the present state. In more precise terms,
 given a
known sequence of ''analogues'', i.e. of past states ${\bf x}_1, ... , {\bf x}_M$
  which resemble each other closely in pairs, so that
$|{\bf x}_k- {\bf x}_M|<\epsilon$    with $\epsilon$ reasonably small, one makes
the approximate prediction:
$$
{\bf x}_{M+1}={\bf x}_{k+1}
$$
if  ${\bf x}_k$  is an analogue of ${\bf x}_M$~\cite{ccv}.

In the case the above protocol can be used,
one may then proceed to build a model of the phenomenon,
i.e. to determine a function ${\bf f}({\bf x})$ such that the sequence of states is well approximated
by the dynamical system 
\begin{equation}
{\bf x}_{k+1}={\bf f}({\bf x}_k) \,\, .
\label{eq:law} 
\end{equation}
 The application of this method requires
knowledge of at least one analogue. It is possible to state\footnote{
This is the essence of Kac's lemma, a well know result of ergodic theory
~\cite{ccv}.}
 that such knowledge
requires sufficiently long sequences, at least of duration of order 
 $ T_R \sim (L/\epsilon)^D$, where $L$ is the typical length scale of the system, and $D$ is the dimension of the attractor\footnote{
 In conservative cases, e.g. Hamiltonian systems,  $D$ is the number of variables involved in the
 dynamics; {if the system is dissipative, $D$ can be a fractional number and is  smaller than the dimension of the phase-space}}.

The exponential growth of $T_R$ as a function of $D$
has a severe  impact  on our ability to make predictions,
and the building of a model for  the evolution law (\ref{eq:law}),
solely relying on previously acquired data. One can say that $D$ larger than 6 renders
the approach described here useless, because it makes it practically impossible to
observe the ''same'' state twice, i.e. within an acceptable accuracy $\epsilon$.

As already stressed in Section 2.2,  the  {\it state 
of the system}, i.e. the variables which describe the phenomenon 
under investigation, is typically not known.
Therefore an unavoidable technical aspect is the determination of the proper state  of the system
from the study of a time series $\{ u_1, u_2, ..., u_M \}$, where $u$ is an observable.
The most relevant result for such a problem is due to Takens who has been
able to show that, at least from a mathematical point of view,
it is possible (if we know that the system is deterministic, described by a finite dimensional vector,
and $M$ is arbitrarily large) to determine
a proper state-variable ${\bf X}$.
In a nutshell: there is a finite integer $m$ such that the delay coordinate
vector (of dimension $m$)
\begin{equation}
{\bf y}_k^{(m)}=(u_k, u_{k-1}, .. , u_{k-m+1})
\end{equation}
can faithfully reconstruct the properties of the underlying
dynamics\footnote{A rigorous result states: $m \ge 2 [D] +1$;
from  heuristic arguments on can expect that  $m= [D]+1$ is enough~\cite{ccv}.
}.

Of course the practical limitation due to the exponential increasing of  $T_R$ as a function of $D$,
is present also in the Takens's method; therefore we have   rather severe practical limitations~\cite{ccv}.
{Indeed, the conceptual idea behind all these inductive approaches is always to try a reconstruction of the relevant phase-space, which, at a resolution level $\epsilon$, has roughly a volume of $ (L/\epsilon)^D$.}
To be explicit, and stress the limit of the method,  consider a system ruled by a deterministic law, for which the dimension of the attractor is
$D$, and  we know  a time series $\{ u_1, u_2, ..., u_M \}$ of  an observable;
the method of Takens allows to find (an approximation of ) the evolution law only if $M$ is larger compared with $A^D$.
The value of  $A$ depends on the wished accuracy; just to give an idea let us assume $A=100$,
corresponding to just a fair accuracy, for $D=6$, $7$, and $8$ we have
$A^D = 10^{12}$, $10^{14}$ and $ 10^{16}$ respectively.
Therefore also in the case  we know that 
  the system is ruled by a deterministic law,
  such a knowledge  does not imply the actual possibility to perform an explicit compression.

\subsection{Discussion}

McAllister\cite{mcallister2003algorithmic} has observed
that empirical data sets are algorithmically incompressible, concluding that the task
of scientific laws and theories does not consist in compressing empirical data. 
{We share such an opinion  on the incompressibility of generic empirical data,
even though his argument is maybe too sharp.
}
As previously discussed  in the context of chaotic deterministic systems (e.g. the Bernoulli's shift), the typical
output is incompressible and,  from a mathematical point of view,
 such a result  is  a consequence of  the important result 
obtained by  Martin-L\"of~\cite{martin1966definition}: almost all the initial conditions correspond to incompressible sequences.

Regarding the opinion that scientific laws constitute a compression of empirical
data, McAllister claims that no scientist has ever made such a statement.
We do not enter into the historical aspects.
However  we want to discuss the following example: consider a series
of light-refraction experiments, in which $\{ \alpha_1, .... , \alpha_N \}$ 
are the angles of the incident
rays, and $\{\beta_1, ... , \beta_N \}$  the angles of the refracted rays. 
The sequences 
$\{ \alpha_1, .... , \alpha_N \}$  and
$\{\beta_1, ... , \beta_N \}$ 
may or may not be compressed.
 This is a frozen accident which
depends on the protocol followed by the scientist while preparing the experiment,
for instance, in the case of the protocol $\alpha_{n+1}=\alpha_n+ \delta$,
 the sequences can be compressed,
 on the contrary if each $\alpha_n$ is selected according to a random rule,
 the sequences are not compressible.
However, once the values  $\{ \alpha_1, .... , \alpha_N \}$  are known, the sequence
 $\{\beta_1, ... , \beta_N \}$ is
simply determined by the Snell's law: $\sin \alpha/ \sin \beta=n$, and this is a genuine form of
compression.
Therefore the example about the Snell's laws, which is often cited 
(likely because mentioned by Mach),
is not particularly deep.

A less trivial instance concerns the Navier-Stokes equation for fluids.
We saw how  in such a  chaotic systems, although
the time sequences are not compressible, the NSE have a predictive
power, in the sense that they are able to generate  results 
in good  qualitative and quantitative agreement with the experiments.

The claim that the world is comprehensible because it is algorithmically compressible
is, in our opinion, a truism, which is equivalent to saying that laws
of nature exist. 
We note that the actual possibility to understand
the world arises mainly  from  a series of lucky facts, in particular:
\\
$\bullet$  Typically physical laws obey spatially and temporally local rules, i.e. a given phenomenon
is not affected too much by events which are distant in time and/or in space.
Practically the main laws of physics, like the  equations of Maxwell, Schrš\"odinger,
Newton etc., obey the locality assumption and are described by differential equations.
\\
$\bullet$  Despite the enormous complexity and the intricate interconnections of different
phenomena, often there is a scale separation which allows us for  a
description   in terms of effective theories
of the different levels on which reality may be considered. 
\\
A celebrated example of an effective theory 
coming from the use  of the  separation of scales which characterises
 the microscopic, and the macroscopic realms,
is the Langevin equation describing the  Brownian motion.

On the other hand, if the laws  are not known and we have just the possibility
to study time series, the scenario is quite pessimistic.
If the effective dimensionality is (relatively) large, 
even in the most simple case of deterministic system, it is not possible to find
the evolution laws and therefore to perform an explicit  compression~\cite{ruelle1990claude}.
Therefore the idea \emph{ law $=$ possibility of compression of data}, must be (re)considered
with many caveats.
{Perhaps we live in a ``big data" era, but actually not big enough to model complex phenomena,
without the help of some theory.}

We stress again that disregarding the distinction between initial conditions and
laws of nature can lead to great confusion. 
In the Introduction we cited Davies\cite{davies1990physical} who claims that chaotic systems
are not algorithmically compressible. The discussion in Sect. 3 shows
how chaotic systems can be trivially compressible (in the sense that it is easy
to write down the evolution laws, as, e.g., for the Bernoulli's shift or the NSE). What
can be not compressible is the output and this is related to the complexity
of the sequence associated to the initial condition.

McAllister after an analysis of the relevance of compressibility in
science concludes\cite{mcallister2003algorithmic}:

{\it  In sum, a scientific law or theory provides an algorithmic compression not of a
data set in its entirety, as Mach, Solomonoff and others believed, but only of a
regularity that constitutes a component of the data set and that the scientist picks
out in the data. The remaining component of the data set, which is algorithmically
incompressible, is regarded as noise in the sense of classical information theory.}

We may agree with the previous sentence, if we keep our eyes open.
To be able to distinguish a regularity (the law) from underlying noise,   a proper resolution and, capitally, the separation of scale which permits 
to build  coarse-graining description are
necessary.
Moreover, in our opinion, for the understanding of any nontrivial topic
it is too naive to hope in an approach based on  data and algorithms,
and  the study of the phenomenal framework is unavoidable\cite{vulpiani2014large}.

Some remarks are in order.
First, we would like to stress
an issue which, although rather important, is often  not discussed.
The relevance of the scale resolution is  closely
linked with the proper effective variables which are able to  describe  the
phenomenon under investigation.
Let us  consider a  fluid which can be described, at microscopic level,
in terms of its molecules; in such an approach the
correct variables are the positions and momenta of the molecules. So we
have a very accurate description containing a lot of informations.
However, the microscopic level  sometimes is not interesting. For instance
in engineering (or geophysical) problems it is much more relevant to
adopt an hydrodynamical description in terms of few fields (velocity, temperature and so on).
Of course using such a macroscopic description
one has a huge decreasing of the amount of information and an increasing of
the possibility to compress data.

Then, few words on the qualitative aspects of science. 
Often qualitative results are considered less important than the quantitative
ones.
That is an unfair view, since although some results cannot be expressed in terms of numerical sequences,  they
can be interesting and rigorous. For instance it can be important to know
that a phenomenon is periodic or some variables are bounded in a certain domain.
We can mention the Lotka-Volterra like equations, for which sometimes one can show that
the time behaviour is periodic, even though it is not possible to  find the explicit solution.
In a similar way, in some celestial mechanics
problems, it is enough to be sure that the motion (e.g. of an asteroid) remains
in a bounded region\cite{ccv}.
The previous qualitative results, although they cannot be formalized in terms
of algorithmic compression (which involves sequences) are genuine forms of
compression of information.

Finally, we would like to indicate an important example of a complex system which has been understood rather successfully thanks to a \emph{traditional} theory/data approach: weather forecasting. 
This problem is related to the dynamics of atmosphere which is characterized by (i) huge degrees of freedom;(ii) dynamical interaction with a complex environment; (iii) chaos and many non-linear feedback mechanisms.
 Nevertheless, the modern developments of  weather forecasting are based on the basic theory of such a system, fluid mechanics as pioneered by Richardson~\cite{chibbaro2014reductionism,vulpiani2014lewis}.
In particular, the use of the theory and the analysis of many phenomenological data have led to the recognition of different separated scales, which has been key for the development of a hierarchy of models adequate at different scales today solved by numerical integration. Those models are statistical and mainly qualitative, but rigorous to some extent thanks to the separation of scale.
This example should point out that the path to the understanding of complex phenomena is a brilliant interplay of deep empirical analysis, creative theoretical developments and technical developments.
\section*{Acknowledgements}
We thank M. Falcioni for his remarks and suggestions and we thank in a special way A. Decoene for her careful reading of the manuscript.

\appendix
\section{The algorithmic complexity in a nutshell}

The rationalization  of the idea of ''randomness'' needs the introduction of 
a precise mathematical formalisation of the complexity of a sequence.

This  has been
proposed independently in 1965 by Kolmogorov, Chaitin and Solomonoff, and refined by Martin-L\"of\cite{li2009introduction,martin1966definition}.

Given the sequence $a_1, a_2, ..., a_N$, 
among all possible programs
which generate this sequence one considers  with the smallest number of instructions.
Denoting by $K(N)$ the number of these instructions, the algorithmic complexity
of the sequence is defined by
$$
K=\lim_{N \to \infty} {K(N) \over N} \,.
$$
Therefore, if there is a simple rule that can be expressed by a few instructions, the
complexity vanishes. If there is no explicit rule, which is not just the complete list
of $0$  and $1$, the complexity is maximal, that is $1$. Intermediate values of $K$ between
$0$ and $1$ correspond to situations with no obvious rules, but such that part of the
information 
necessary to do a given step is contained in the previous steps.

To give an intuitive idea of the concept of complexity, let us consider a situation
related to the transmission of messages\cite{chaitin1990information}: A friend on Mars needs the
tables of logarithms. It is easy to send him the tables in binary language; this method
is safe but would naturally be very expensive. It is cheaper to send the instructions
necessary to implement the algorithm which computes logarithms:
it is enough to specify few simple properties, e.g.
$$
\ln (a \, b)=\ln (a) + \ln (b) \,\, , \,\, \ln (a^{\alpha} b^{\beta}) = \alpha \ln (a) + \beta \ln (b) \,\, ,
$$
and, in addition, for $|x|<1$ the following Taylor expansion:
$$
\ln (1+x)=\sum_{n=1}^{\infty}(-1)^{n+1} {x^n \over n} \,\,.
$$
However, if the friend is not interested in mathematics, but rather in football or
the lottery, and wants to be informed of the results of football matches or lottery
draw, there is no way of compressing the information in terms of an algorithm
whose repeated use produces the relevant information for the different events; the
only option is the transmission of the entire information.
To sum up: the cost of the transmission of the information contained in the
algorithm of logarithms is independent of the number of logarithms one wishes
to compute. On the contrary, the cost of the transmission of football or lottery results
increases linearly with the number of events. One might think that the difference is
that there are precise mathematical rules for logarithms, but not for football matches
and lottery drawings, which are then classified as random events.

\bibliographystyle{unsrt}
\bibliography{biblio-1}

\begin{thebibliography}{10}

\bibitem{seve3}
L.~S\`eve.
\newblock {\em Penser avec Marx aujourd'hui: philosophie ?}
\newblock La Dispute, 2014.

\bibitem{bailly2006mathematiques}
Francis Bailly and Giuseppe Longo.
\newblock {\em Math{\'e}matiques et sciences de la nature}.
\newblock Hermann, 2006.

\bibitem{Mach1}
Ernst Mach.
\newblock {\em On the Economical Nature of Physical Inquiry}.
\newblock Cambridge University Press, 2014.

\bibitem{Mach2}
Ernst Mach.
\newblock {\em The science of mechanics: A critical and historical account of
  its development}.
\newblock Open court publishing Company, 1907.

\bibitem{li2009introduction}
Ming Li and Paul~MB Vit{\'a}nyi.
\newblock {\em An introduction to {K}olmogorov complexity and its
  applications}.
\newblock Springer Science \& Business Media, 2009.

\bibitem{solomonoff1964formal}
Ray~J Solomonoff.
\newblock A formal theory of inductive inference. part i-ii.
\newblock {\em Information and control}, 7(1):1, 1964.

\bibitem{davies1990physical}
Paul~CW Davies.
\newblock Why is the physical world so comprehensible.
\newblock {\em Complexity, entropy and the physics of information}, pages
  61--70, 1990.

\bibitem{barrow2007new}
John~D Barrow.
\newblock {\em New theories of everything}.
\newblock Oxford University Press, 2007.

\bibitem{born1949natural}
Max Born.
\newblock {\em Natural philosophy of cause and chance}.
\newblock Read Books, Vancouver, 1948.

\bibitem{Sok_97}
A.D. Sokal and J.~Bricmont.
\newblock {\em Postmodern Intellectuals}.
\newblock Picador, 1997.

\bibitem{ccv}
Massimo Cencini, Fabio Cecconi, and Angelo Vulpiani.
\newblock {\em Chaos}.
\newblock World Scientific, 2010.

\bibitem{coveney2016big}
Peter~V Coveney, Edward~R Dougherty, and Roger~R Highfield.
\newblock Big data need big theory too.
\newblock {\em Phil. Trans. R. Soc. A}, 374(2080):20160153, 2016.

\bibitem{crutchfield2014dreams}
James~P Crutchfield.
\newblock The dreams of theory.
\newblock {\em Wiley Interdisciplinary Reviews: Computational Statistics},
  6(2):75--79, 2014.

\bibitem{dyson2009birds}
Freeman Dyson.
\newblock Birds and frogs.
\newblock {\em Notices of the AMS}, 56(2):212--223, 2009.

\bibitem{chibbaro2014reductionism}
Sergio Chibbaro, Lamberto Rondoni, and Angelo Vulpiani.
\newblock {\em Reductionism, Emergence and Levels of Reality}.
\newblock Springer, 2014.

\bibitem{times}
In Neil A.~Gershenfeld Andreas S.~Weigend, editor, {\em Time series prediction:
  Forecasting the future and understanding the past}. Addison-Wesley, Reading,
  1994.

\bibitem{onsager1953fluctuations}
Lars Onsager and S~Machlup.
\newblock Fluctuations and irreversible processes.
\newblock {\em Physical Review}, 91(6):1505, 1953.

\bibitem{Ma}
S.K. Ma.
\newblock {\em Statistical Mechanics}.
\newblock World Scientific, 1985.

\bibitem{takens1981detecting}
Floris Takens.
\newblock Detecting strange attractors in turbulence.
\newblock In D.~A. Rand and L.~S. Young, editors, {\em Dynamical Systems and
  Turbulence}. Springer, Berlin, 1981.

\bibitem{eckmann1992fundamental}
J-P Eckmann and David Ruelle.
\newblock Fundamental limitations for estimating dimensions and lyapunov
  exponents in dynamical systems.
\newblock {\em Physica D: Nonlinear Phenomena}, 56(2):185--187, 1992.

\bibitem{newton1934sir}
Isaac Newton.
\newblock {\em Sir Isaac Newton's mathematical principles of natural philosophy
  and his system of the world}.
\newblock Univ of California Press, 1934.

\bibitem{wigner1963problem}
Eugene~P Wigner.
\newblock Events, laws of nature and invariant principles.
\newblock {\em Science}, 145:995, 1964.

\bibitem{li1992inductive}
Ming Li and Paul~MB Vit{\'a}nyi.
\newblock Inductive reasoning and {K}olmogorov complexity.
\newblock {\em Journal of Computer and System Sciences}, 44(2):343--384, 1992.

\bibitem{li1994kolmogorov}
Ming Li and Paul~MB Vit{\'a}nyi.
\newblock Kolmogorov complexity arguments in combinatorics.
\newblock {\em Journal of Combinatorial Theory, Series A}, 66(2):226--236,
  1994.

\bibitem{ming2014kolmogorov}
Ming Li and Paul~MB Vit{\'a}nyi.
\newblock Kolmogorov complexity and its applications.
\newblock {\em Algorithms and Complexity}, page 187, 2014.

\bibitem{martin1966definition}
Per Martin-L{\"o}f.
\newblock The definition of random sequences.
\newblock {\em Information and control}, 9(6):602--619, 1966.

\bibitem{caludelongo}
Cristian Calude and Giuseppe Longo.
\newblock Classical, quantum and biological randomness as relative
  unpredictability.
\newblock {\em Natural Computing}, 15:263, 2016.

\bibitem{dirac1929quantum}
Paul Adrien~Maurice Dirac.
\newblock Quantum mechanics of many-electron systems.
\newblock {\em Proc. Roy. Soc. London. Ser. A}, 123(792):714--733, 1929.

\bibitem{Pri_81}
H.~Primas.
\newblock {\em Chemistry, Quantum Mechanics and Reductionism. Perspectives in
  Theoretical Chemistry}.
\newblock Springer, Berlin, 1981.

\bibitem{Scerri2008}
E.R. Scerri.
\newblock {\em Collected papers on philosophy of chemistry}.
\newblock World Scientific, 2008.

\bibitem{Jon_10}
G.~Jona-Lasinio.
\newblock Spontaneous symmetry breaking -- variations on a theme.
\newblock {\em Progress of Theretical Physics}, 124(5):731, 2010.

\bibitem{frisch1995turbulence}
Uriel Frisch.
\newblock {\em Turbulence: the legacy of AN Kolmogorov}.
\newblock Cambridge university press, 1995.

\bibitem{bohr2005dynamical}
Tomas Bohr, Mogens~H Jensen, Giovanni Paladin, and Angelo Vulpiani.
\newblock {\em Dynamical systems approach to turbulence}.
\newblock Cambridge University Press, 2005.

\bibitem{Kan_97}
H.~Kantz and T.~Schreiber.
\newblock {\em Nonlinear time series analysis}.
\newblock Cambridge University Press, 1997.

\bibitem{mcallister2003algorithmic}
James~W McAllister.
\newblock Algorithmic randomness in empirical data.
\newblock {\em Studies in History and Philosophy of Science Part A},
  34(3):633--646, 2003.

\bibitem{ruelle1990claude}
David Ruelle.
\newblock The {C}laude {B}ernard lecture, 1989. deterministic chaos: the
  science and the fiction.
\newblock {\em Proceedings of the Royal Society of London A: Mathematical,
  Physical and Engineering Sciences}, 427(1873):241--248, 1990.

\bibitem{vulpiani2014large}
In Angelo Vulpiani, Fabio Cecconi, Massimo Cencini, Andrea Puglisi, and Davide
  Vergni, editors, {\em Large Deviations in Physics: The Legacy of the Law of
  Large Numbers}, volume 885. Springer, 2014.

\bibitem{vulpiani2014lewis}
Angelo Vulpiani.
\newblock Lewis {F}ry {R}ichardson: scientist, visionary and pacifist.
\newblock {\em Lettera Matematica}, 2(3):121--128, 2014.

\bibitem{chaitin1990information}
Gregory~J Chaitin.
\newblock {\em Information, randomness \& incompleteness: papers on algorithmic
  information theory}.
\newblock World Scientific, 1990.

\end{thebibliography}

\end{document}